\documentclass[twocolumn,showpacs,preprintnumbers,amsmath,amssymb]{revtex4}
\usepackage{graphicx}
\usepackage{dcolumn}
\usepackage{bm}

\begin{document}
\title{Parameter fluctuations in coupled chaotic systems}
\author{M. P. John}
\email{manupj@cusat.ac.in}
\author{P.U. Jijo}
\email{jijo@photonics.cusat.edu}
\author{V. M. Nandakumaran}
%\email{nandak@cusat.ac.in}
\homepage{http://www.photonics.cusat.edu}
\affiliation{International
School of Photonics, Cochin University of Science and
Technology.\\Cochin 682 022, India }
\date{\today}
\begin{abstract}
We study the effect of parameter fluctuations on synchronization of
a coupled chaotic system. The fluctuations to the parameter can be
random or it can be a periodic modulation. For random fluctuations
we introduce a new quantity, the \emph{rate of fluctuations}, apart
from the statistical features of the fluctuations. Fluctuation rate
in our study refers to the number of random  modifications to
parameters occurring in unit time. With a periodic modulation, the
fluctuation rates can be the frequency of the modulating term. It is
found that with high fluctuation rates the synchronization is stable
irrespective of the statistical or mathematical features of
fluctuation. It was also found that the low fluctuation rates can
destroy synchronization even with a small amplitude. We analytically
explain the observed phenomenon using the dynamical equations and
numerically verified with a coupled system of Rossler attractor as
an example. We also numerically quantify the relation between
synchronization error and fluctuation rates.
\end{abstract}

 \maketitle

\section{Introduction}
Synchronization of coupled chaotic systems has generated a lot of
research activities over the last several years. Synchronized
behavior has been studied extensively in physical, chemical and
biological systems.Different types of synchronization such as
complete, generalized, lag and phase synchrony are described in
literature. One of the methods by which the synchronization of
chaotic systems is achieved is by coupling two identical systems,
which may be unidirectional or bidirectional\cite{yamada, peco2,
peco3,everyone, war1, war2, roy2, bindu,kurth}. Synchronization in
arrays of coupled laser systems has also been investigated under
various coupling schemes\cite{everyone, war1, war2, roy2, bindu}.
Complete synchronization of identical chaotic systems is also of
considerable interest because of its applications in secure
communication\cite{war2, bindu}. By identical systems we mean a set
of systems whose parameters are exactly equal. It is found that the
synchronization is not robust when there is a small but finite
mismatch of the parameters of the systems\cite{ros,pslag,kurth}. In
fact the effect of phase mismatch and a finite constant frequency
detuning in a bidirectionally coupled Duffing oscillators is to
destroy the synchronization altogether\cite{yin}.

In the present paper we address a different issue that is relevant
in many practical physical systems. We study the effect of random
fluctuations in the parameters of the system on the synchronization
properties. Such a study is relevant and important since in a real
physical system the parameters can often fluctuate randomly either
due to some internal instabilities or due to some external
perturbations.

The paper is organized as follows; in section II. we consider random
perturbations to one of the parameter that characterizes the
synchronization. We present the criteria for the robustness of
synchronization. In section III. the numerical results on two
coupled Rossler systems with randomly fluctuating parameters are
presented. Section IV contains the discussions of the result.

\section{Parameter Fluctuations in Coupled Systems}

In this section we consider two identical dynamical systems which
are coupled together. To study the effect of fluctuations it is
essential to identify one parameter whose mismatch is most effective
in destroying synchronization. We denote this parameter as $p$. Then
the equations for the coupled systems are given by.

\begin{eqnarray}
\label{one}
\dot{X_1}&=&f_1(p_1, X_1)+C f(X_2-X_1)\\
\dot{X_2}&=&f_1(p_2, X_2)+C f(X_1-X_2)\nonumber
\end{eqnarray}

Here $C$ is the coupling constant. In reality it is difficult if not
impossible to construct identical systems except in numerical
simulations. This can also be due to the fact that the parameters
could be fluctuating in time with a time scale of their own. We
incorporate this by writing the parameter as

 \begin{eqnarray}\label{addfluct}
 p_1 = p_0+\xi_{1t}\\
 p_2 = p_0+\xi_{2t},\nonumber
 \end{eqnarray}
where, $\xi_{1t}$ and $\xi_{2t}$ are two delta correlated zero mean
Gaussian random variables. A measure of the amplitude of
fluctuations, we define $\widetilde{\Delta p}$, as
\begin{equation}
\widetilde{\Delta p}=\langle \mid \delta p (t) \mid \rangle_t,
\end{equation}
where, $ \delta p(t) = p_1(t)-p_2(t)$ and $\langle ...\rangle_t$
denotes time average.

To study the effect of time scales of parameter fluctuation, we
define the fluctuation rate $\phi= number~ of~ perturbuations /
unit~time$. Different fluctuation rates can be achieved numerically
by modifying the parameter as in Eqn\ref{addfluct} only in certain
chosen time steps. Rest of the time the value of the parameter
remains constant at the modified value. The Error in synchrony is
studied varying $\phi$. The effect of time scales has not been
studied in literature and our results indicate that it is highly
significant in determining the quality of synchronization.

We also considered a periodic modulation of the parameters, where
the Eqn. \ref{addfluct} can be replaced by

\begin{eqnarray}
\label{amodf}
p_{1}&=&p_0+a \sin{ft}\\
p_{2}&=& p_0-a \sin{ft}\nonumber
\end{eqnarray}

where $f$ is the frequency  and $a$ is the amplitude of modulation.
By choosing an appropriate value for $a$ and by changing $f$ the
quality of synchronization for various modulating frequencies can be
studied.

\section{Numerical examples}
Coupled Rossler oscillators are a well known for numerical studies
in synchronization. Consider a system of bidirectionally coupled
Rossler oscillators. The coupled equation can be written as.
\begin{eqnarray}
\dot{x_1}&=&- y_1 - z_1+c(x_2-x_1)\\\nonumber \dot{y_1}&=& x_1 +
p_1y_1\\\nonumber \dot{z_1}&=& 0.2 + z_1 ( x_1 - 10 )\\\nonumber
\dot{x_2}&=&- y_2 - z_2+c(x_1-x_2)\\\nonumber
\dot{y_2}&=& x_2 + p_2y_2\\
\nonumber \dot{z_2}&=& 0.2 + z_2 ( x_2 - 10 )\\\nonumber
\end{eqnarray}

\begin{figure}[tbh]
\centering
\includegraphics[width=0.9\columnwidth]{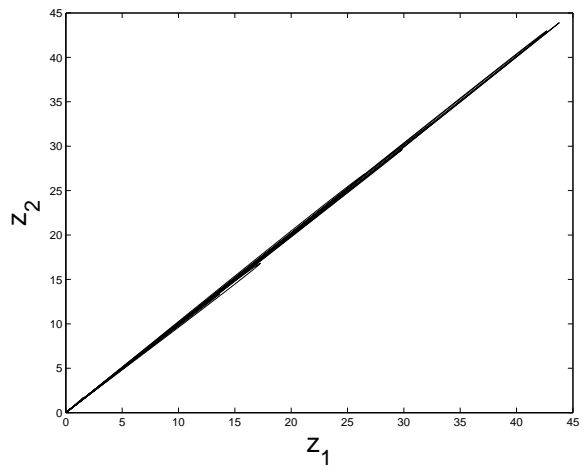}
\caption{Synchronization is maintained in the presence of parameter
fluctuations. $\phi~=~1000$ and $\widetilde{\Delta p} =0.05$}
\label{hifluct}
\end{figure}

Here, the coupling strength $c=0.15$, and $p_0 = 0.18$. Though the
coupling strength can also affect synchronization, we chose a value
that is best suited for illustrating the concepts. Also the value of
 $\widetilde{\Delta p}$ was fixed to be 0.05 for all fluctuation rates.

\begin{figure}[tbh]
\centering
\includegraphics[width=0.9\columnwidth]{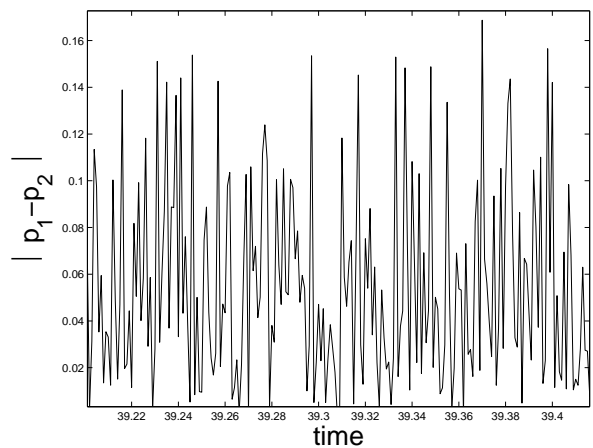}
\caption{Synchronized system also posses a zero mean parameter
mismatch. } \label{mismatch}
\end{figure}

\begin{figure}[tbh]
\centering
\includegraphics[width=0.9\columnwidth]{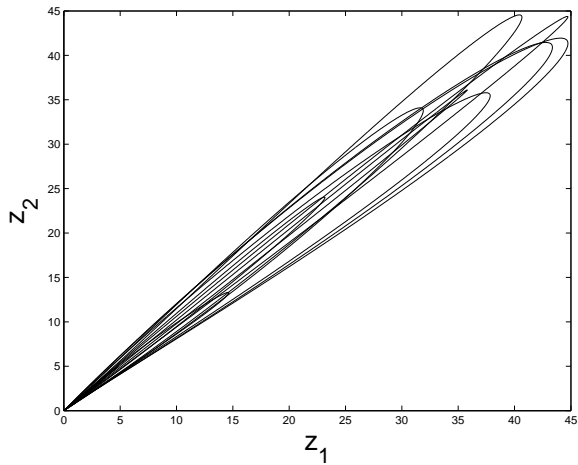}
\caption{Synchronization is destroyed in the presence of parameter
fluctuations with low fluctuation rates. Fluctuation rate
$\phi~=~25$ and $\widetilde{\Delta p} =0.05$} \label{lofluct}
\end{figure}

 Fig.\ref{hifluct} shows the synchronization plot in the presence of
parameter fluctuations. It can be seen that the synchronization is
robust. Fig.\ref{mismatch} shows that the coupled systems posses a
parameter mismatch at any instant of evolution of the system. Also
at times the instantaneous mismatches can be compared to the value
of the average value parameter value itself. With the same value of
$\widetilde{\Delta p}$ the synchronization is destroyed with a lower
fluctuation rate as shown in Fig.\ref{lofluct}.

We have studied the relationship between the synchronization error
and the fluctuation rate. To quantify the synchronization error we
used the similarity function given by

\begin{equation}\label{dynamic1}
S^2(\tau)=\frac{\langle[x(t+\tau)-x(t)]^2\rangle}{[\langle
x^2(t)\rangle \langle x^2(t)\rangle]^{\frac{1}{2}}}.
\end{equation}

\begin{figure}[tbh] \centering
\includegraphics[width=0.9\columnwidth]{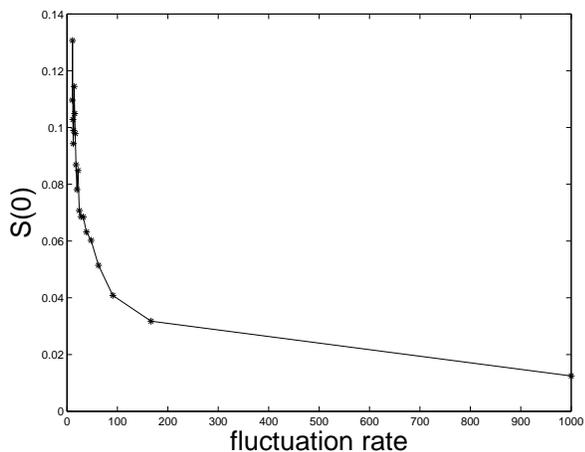}
\caption{The synchronization error decreases with the increase in
the fluctuation rate.} \label{e_rate}
\end{figure}

Here $\tau$ is set to zero, which gives $S(0)$, the error in
synchrony. Fig. \ref{e_rate} Shows the plot of $S(0)$ vs.
fluctuation rate. It can be seen the error diminishes rapidly with
the increase in the fluctuation rate.

\begin{figure}[tbh]
\centering
\includegraphics[width=0.95\columnwidth]{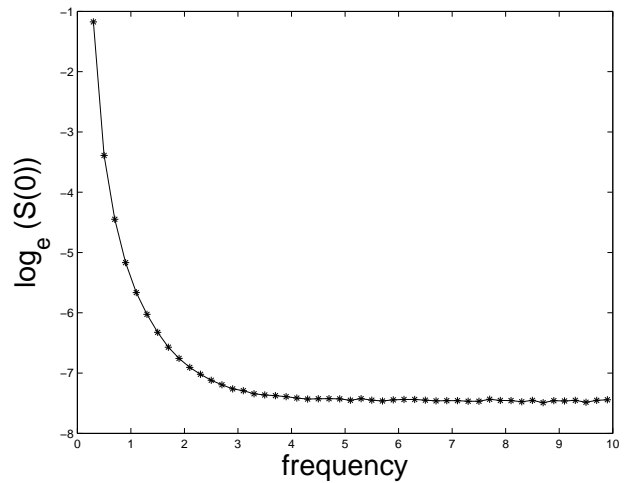}
\caption{The synchronization error decreases with the increase the
modulating frequency.} \label{e_freq}
\end{figure}

A similar behavior was also seen with the periodic modulation of
parameters. Fig. \ref{e_freq} shows that the synchronization error
decreases with increase in the modulating frequency. This clearly
suggests that the fluctuation frequency or rate is more important
than the nature of fluctuations.

\section{Discussion}

The robustness of synchronization with high fluctuation rates and
destruction of synchronization with low fluctuation rates can be
explained analytically as follows. With Eqn.\ref{one} we can write
an equation for the rate of separation $ X_1- X_2 $ of the
trajectories as,
\begin{equation}
\frac{d (X_1- X_2)}{dt}= \dot{X_1}- \dot{X_2}=M(p_1, p_2, X_1,X_2),
\end{equation}
 $M(p_1, p_2, X_1,X_2)$ is a function of the dynamical
variables and the parameters of the coupled systems. This can be
written as the sum of two terms,
\begin{equation}
M(X_1,X_2)=S(X_1,X_2)+E(X_1,X_2).
\end{equation}

This comes from the fact that for a function $\Omega (p,X)$, we can
write for small $\Delta p$ and  neglecting its higher powers or if
the higher derivatives of $\Omega$ w.r.t $p$ is zero,

\begin{equation}
\Omega (p+\Delta p,X)= \Omega (p,X)+ \Delta p\;\frac{\partial \Omega
(p,X)}{\partial p}.
\end{equation}

This can be valid for functions in the dynamical equations of a
chaotic system if the parameter values are not near a bifurcation
point. Using this relation, with $p_1=p_0+\Delta p_1$ and
$p_2=p_0+\Delta p_2$,
\begin{eqnarray*}
S(p_0,X_1,X_2)&=& F_1(p_0,X_1, X_2)-F_2(p_0, X_1, X_2)\\
\end{eqnarray*}

and

\begin{eqnarray*}
E(p_1,p_2,X_1,X_2)&=&  \Delta p_1 \frac{ \partial F_1(p_1,p_2,X_1,
X_2)}{\partial
p}\mid_{p_1=p_0}\\
& & - \Delta p_2 \frac{\partial F_2(p_1,p_2,X_1,
X_2)}{\partial p}\mid_{p_2=p_0}.\\
\end{eqnarray*}
with $\Delta p_1~=~\xi_{1t}$ and $\Delta p_2~=~\xi_{1t}$.

Here $S(p_0,X_1,X_2)$ represents the quantity which offers a stable
synchronization manifold, that is, when $S(p_1,p_2,X_1,X_2)$ alone
in the right hand side of the separation equation, coupled systems
synchronize as $t \longrightarrow \infty$. The conditions for such a
synchronization is widely discussed in literature \cite{funda}. The
term $E(p_1,p_2,X_1,X_2)$ represents the effect of the parameter
mismatch. Coupled systems can synchronize if the overall effect of
this term is zero as $t \longrightarrow \infty$. One possible way
for this is when $E(p_1, p_2,X_1,X_2,C)$ is of the form

\begin{eqnarray}\label{vanish}
E(p_1, p_2,X_1,X_2,C)= \sum_i \rho(t)_ix_i(t)
\end{eqnarray}

where $\rho_i (t)$ is the the fluctuation term and $x(t)$ is the
phase space variables of the coupled system. Here the equation
vanishes because the $\rho (t)$'s are zero mean rapidly fluctuating
quantities and $x(t)$'s are the phase space variables that evolve
slowly when compared to the the rapid fluctuations of the parameter.
Thus $x(t)$'s can be assumed to be constant, in the time required
for the fluctuations get summed to zero. This also explains why the
synchronization is destroyed when the fluctuation rate is low. With
a low fluctuation rate the $E(p_1, p_2,X_1,X_2,C)$ cannot be summed
to zero every time since the phase space evolution time is
comparable to the interval where a fixed parameter mismatch
persists. Thus with a lower fluctuation rate the system always get
time to respond to the parameter mismatch before it being canceled
out. Also with a slowly varying parameter mismatch, a definite state
of phase synchrony is also not attained and the system remains in a
transient state through out the evolution in the phase space.

In the present example, it can be seen that the quantity $E(p_1,
p_2,X_1,X_2,C)$ can be expressed as,

\begin{eqnarray}
E(p_1, p_2,X_1,X_2,C)&=& \xi_{1t}y_1-\xi_{2t}y_2
\end{eqnarray}
because the fluctuating terms appear only in the equation of
$\dot{y}$ only. Similar studies were done to the other parameters of
the coupled systems as well, which gave similar result. Apart from
gaussian random fluctuations, we studied perturbations with a
uniform distribution. The results were qualitatively the same as for
the gaussian perturbations which suggests that the most important
quantity that determines the stability of synchronization is the
fluctuation rates.

\section{CONCLUSIONS}
In this paper we studied the effect of parameter fluctuations on the
synchronization of  coupled chaotic systems. We investigated random
parameter fluctuation and also a periodic modulation to the
parameter. It was found that the most significant entity that
determines the quality of synchronization is the fluctuation rates
that we have defined or the frequency of fluctuation. Our study also
show that the timescales with which the parameter fluctuates is more
significant than the statistical or mathematical features of the
fluctuations.

The effect of noise on synchronization has been studied in the past.
Noise affects synchronization in different manner in various
situations. In most of the cases noise destroys synchronization or
make it unfit for the secure communication purposes that we have
cited in the introduction \cite{ashwin,gauth}. Also there are cases
where synchronization is robust to noise\cite{carollnew} or even
induce synchronization \cite{analytical}.

The effect of noise and parameter fluctuations are different. Noise
induces perturbations to the phase space variables that decay while
the system evolves. In a case where the parameter fluctuates, the
resultant perturbations do not die out with the evolution of the
system. It remains the same until it is corrected manually or the
fluctuation modifies the parameter to a new value. Due to this
reason, the fluctuation rate plays an important role in determining
the stability of synchronization in coupled chaotic systems.
Parameter fluctuations may also have much higher significance in
coupled arrays of nonlinear oscillators, especially in biological
systems which exhibit synchronized behavior. Though we have not
included these in our present paper, we hope that our studies will
be a motivation in this direction.

\section{ACKNOWLEDGMENTS}

We gratefully acknowledge fruitful discussions of this work with
S.Rajesh. First two authors are supported by the Council for
Scientific and Industrial Research (CSIR), New Delhi.

\bibliographystyle{apsrev}

\end{document}